\documentclass[showpacs,floats,prb,aps]{revtex4}
\usepackage{graphicx}
\usepackage{epsfig}
\begin{document}

\preprint{}
\title{ Spin-dependent Scattering by a Potential Barrier on a Nanotube}
\author{Yonatan Abranyos$^1$}
\email{yabranyo@hunter.cuny.edu}
\author{Godfrey Gumbs$^{1,3}$}
\email{ggumbs@hunter.cuny.edu}
\author{Paula Fekete$^2$}
\email{paula.fekete@usma.edu}

\affiliation{$^1$Department of Physics and Astronomy,
Hunter College at the City University of New York, \\
695 Park Avenue New York, NY 10065}
\affiliation{$^2$United States Military Academy, West Point, New York}
\affiliation{$^3$Donostia International Physics Center (DIPC),
P. de Manuel Lardizabal, 4, 20018 San Sebasti\'an,
Basque Country, Spain }

\date{\today}

%%%%%%%%%%%%%%%%%%%%%%%%%%%%%%%%%%%%%%%%%%%%%%%%%%%%%%%%%%%%%%%

\begin{abstract}
The electron spin effects on the surface of a nanotube have been
considered through the spin-orbit interaction (SOI), arising from
the electron confinement on the surface of the nanotube.
This is of the same nature as the Rashba-Bychkov SOI at a semiconductor
heterojunction. We  estimate the effect of disorder
within  a potential barrier on the transmission probability. 
Using a continuum model, we obtained analytic
expressions for the spin-split energy bands for
electrons on the surface of nanotubes in the presence of SOI.
First we calculate analytically the scattering amplitudes from a potential 
barrier located around the axis of the nanotube into spin-dependent states. 
The effect of disorder on the scattering process is included
phenomenologically and induces a reduction in the transition probability.
We  analyzed the relative role of SOI and disorder on the transmission probability
which depends on the angular and linear momentum of the incoming particle,
and its spin orientation. We demonstrated that in the presence of disorder 
perfect transmission may not be achieved for finite barrier heights. 

\end{abstract}


%%%%%%%%%%%%%%%%%%%%%%%%%%%%%%%%%%%%%%%%%%%%%%%%%%%%%%%%%%%%

\maketitle\section{Introduction}
\label{sec1}

Carbon nanotubes  which are members of the fullerene  family,  have
novel properties that make them potentially useful in many applications
in electronics, optics, and other fields of  nanotechnology and materials
science. Their strength is extraordinary, they possess  unique
electrical properties, and are efficient thermal conductors. The
nanotube diameter is on the order of a few nanometers and the nanotube length
several millimeters. Nanotube ends  might be capped
with a hemisphere of the buckyball structure or some other material
providing a potential barrier for the carrier electrons on the
nanotube. Furthermore, there  may be a coupling between a particle's
intrinsic (spin) and its extrinsic (orbital motion) degrees of freedom,
thereby giving rise to a  spin-orbit interaction (SOI)
term in  the Hamiltonian describing the energy eigenstates of the nanotube.
The SOI may be due to the electromagnetic interaction between the electron's
spin and the nucleus' electric field through which the electron moves.\cite{nitta}
As a matter of fact, the SOI on the nanotube is of the same type as the Rashba
SOI at a hetrojunction of two types of semiconductors which has given rise
to many interesting properties such as spontaneous spin effects in a 
two-dimensional electron gas (2DEG).\cite{gumbs}

Detailed theoretical studies of the spin-orbit coupling in graphene and carbon
nanotubes have been carried out recently.\cite{DeMartino+Egger,guinea,kane,prb1}
This work was done in conjunction with recent experiments which have shown that
a gate voltage applied perpendicular to the axis of a carbon nanotube
can cause an observable SOI.\cite{9a,9,10} These studies were stimulated by the
consequences of the presence of SOI on transport properties in carbon nanotubes. The origin of SOI is that,
when electrons moving in an electrostatic potential  $\Phi({\bf r})$
(due to ions or gate fields) experience an effective magnetic field,
${\bf B}_{eff} \propto {\bf v}\times\nabla \Phi$ in their rest. For graphene, intrinsic SOI comes from next-neighbor
interactions and is therefore small ($\sim\ 10 $ mK= $ 10^{-3}$ meV).
\cite{guinea,prb1} However, nearest-neighbor terms are nonzero when external
(Rashba) electric fields or curvature-induced
overlap changes break the symmetry which is responsible for the vanishing of the
nearest-neighbor hopping. Therefore, dominant SOI in the absence of external
fields is naturally present in carbon nanotubes.
For nanotubes,  the Rashba field gives a  small effect since it is
averaged over the circumference.   \cite{DeMartino+Egger,guinea,kane,prb1}
However, when electrostatic gates are applied to carbon nanotubes, and
in the vicinity of a van Hove singularity,\cite{yonatan} the Rashba SOI is enhanced
and could lead to  observable  effects.

\begin{figure}[p]
\begin{center}
\epsfig{file=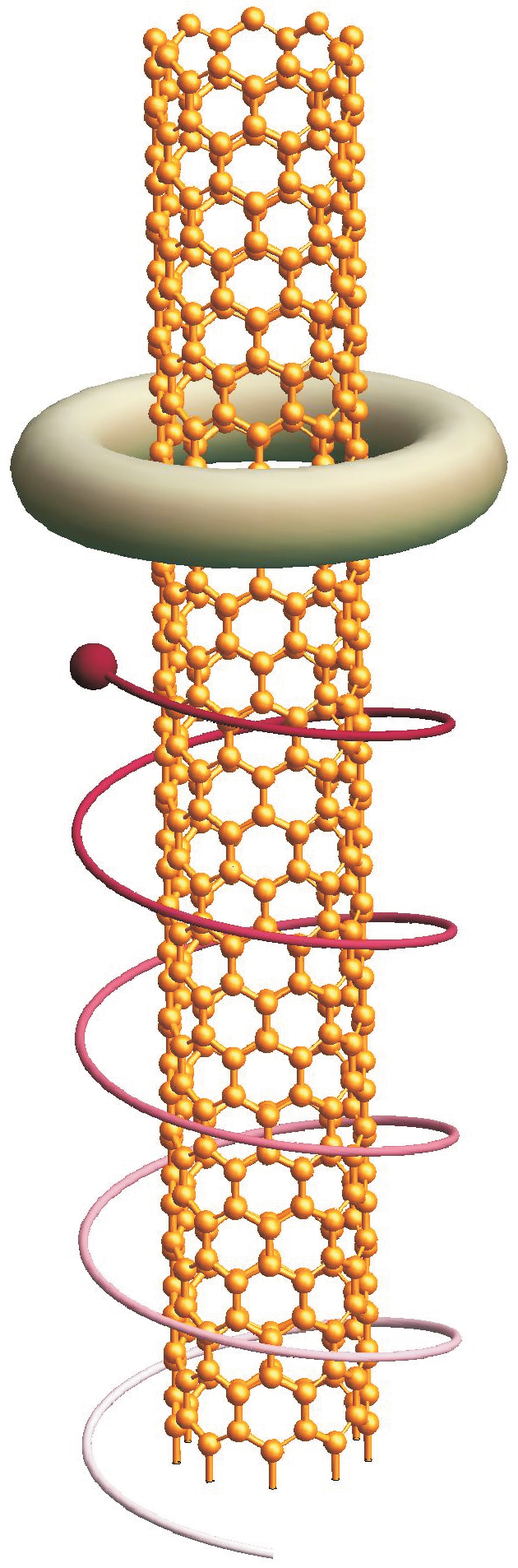,width=4.5in,height=3.8in}
\caption{(Color online) Schematic illustration of the nanotube in the presence
of a particle having linear momentum parallel to the axis of the nanotube
as well as angular momentum around its axis. The incoming particle impinges on a
barrier of uniform thickness.}
\label{FIG:1}
\end{center}
\end{figure}

We will assume that the nanotube we are dealing with is of high quality but will
investigate the role played by disorder within the potential barrier on
the electron transport. The electron eigenstates are  labeled by quantum
numbers representing its two degrees of freedom.\cite{yonatan,yonatan2}   These
are the wave number $k_z$ along the axis of the nanotube and the angular
momentum quantum number $l$ around its axis. The SOI shifts and splits the
energy levels giving eigenstates which may be interpreted as a superposition of
the $|\uparrow\rangle$ and $|\downarrow\rangle$ spin states.  With this background
information, we are motivated to consider a problem involving
spin currents\cite{spincurrent} along the surface of a nanotube and  the
associated transmission and reflection probabilities from a potential barrier.
The potential barrier may be produced by an electrostatic  gate voltage, whose
electric field has the ability  to act as an extra control parameter to exclude
electrons from a region of the nanotube, thereby producing a potential barrier.
Multiple gates (three top gates and a back gate) have been reported to be
deposited on the surface of carbon  nanotubes to produce quantum dots.\cite{marcus}
Sharma\cite{science} has reviewed a method for generating spin currents
and discusses ways in which to overcome some obstacles.
There have been several recent papers dealing with the penetration of
spin-currents through a potential barrier in a
2DEG. \cite{sablikov,tkach,sukhanov,sablikov2} In this regard, we consider an
electron with linear wave number $k_z^i$ and angular momentum quantum number $L^i$
having probability  amplitude $a_+$ in the $|\uparrow\rangle$ state and probability
amplitude $a_-$ in the $|\downarrow\rangle$ state.  The electron is incident on a
cylindrical potential barrier, as shown in Fig.\ \ref{FIG:1}.
We consider the effect of disorder arising from impurities, defects, or by 
the atoms/molecules composing the nanotube which simply oscillate around their equilibrium positions has on the scattering process. We first calculate the 
tunneling  and reflection probabilities in the absence of disorder.  In principle,
these calculations require a knowledge of the electron eigenstates in the presence
of  SOI. For the ballistic transport, we assume
that the electron mean free path is much longer than the diameter  of the
nanotube so that the electron's motion is only altered by interaction with
the potential barrier. We then outline and employ a phenomenological theory to analyze the effect
due to disorder on the tunneling.

The outline of the remainder of this paper is as follows. In Section \ref{sec2},
we present the model spin-orbit Hamiltonian  for single electrons and the
corresponding single-particle eigenstates for an intercalated carbon
nanotube.  Section \ref{sec3} is devoted to calculating the transmission
and reflection probability amplitudes through a  cylindrical potential barrier
encircling the axis of the nanotube in the absence of disorder. In Sec. \ref{newsec4a}, we
introduce and  discuss procedures for including the effect of disorder on the
tunneling probability through the potential barrier.
We present and discuss our numerical calculations in Section \  \ref{sec4}. Some
concluding remarks are given in Section\ \ref{sec5}.

%%%%%%%%%%%%%%%%%%%%%%%%%%%%%%%%%%%%%%%%%%%%%%%%%%%%%%%%%%%%%%%%%%%%%%%%%%%%%%%%%%

\maketitle\section{ The Model Spin-Orbit Hamiltonian}
\label{sec2}

Let us consider a nanotube with its axis along the $z$-axis
in the presence of SO coupling. The Hamiltonian for an electron
with momentum ${\bf p}$ on the surface of a cylinder takes
the following form \cite{yonatan,yonatan2}

\begin{eqnarray}
&&H= \frac{1}{2m^\ast }\left(\hat p^{2}_{z}+\hat p^{2}_{\phi}\right)
+\frac{\alpha_{R}}{\hbar}\left[(\sigma_{1}\sin\phi-\sigma_{2}\cos\phi)\hat p_z
+\sigma_{3}\hat p_{\phi} \right] \ ,
\end{eqnarray}
where $\alpha_R$ is the Rashba SOI parameter  due to radial confinement
and $\sigma_{i}$ are the Pauli matrices. The general solutions are \cite{yonatan}

\begin{eqnarray}
|u_{\nu}(\phi,z)\rangle&=&
\left(
\begin{array}{ccc}
u^{+}_{\nu}(\phi,z) &\\
u^{-}_{\nu}(\phi,z) &
\end{array}\right)
\nonumber\\
u^{(\pm)}_{\nu}(\phi,z)&=&\frac{e^{ikz}}{\sqrt{L_z}}\Phi^{(\pm)}(\phi),\;\;\;\;
\Phi^{(\pm)}(\phi)=\frac{1}{\sqrt{2\pi}}\sum^{\infty}_{l=-\infty}c^{(\pm)}_{l}
(k_z)e^{il\phi}\ .
\end{eqnarray}

\begin{figure}[p]
\begin{center}
\includegraphics[width=3.6in]{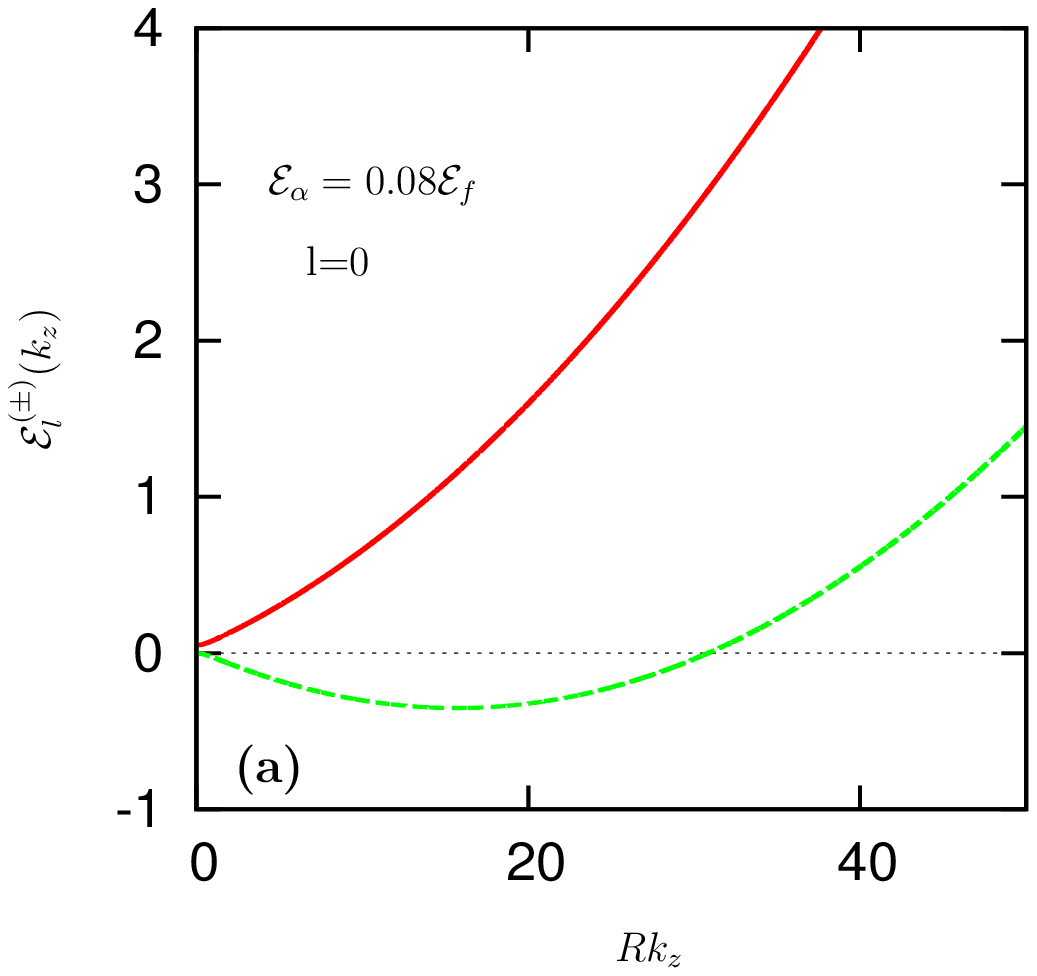}
\includegraphics[width=3.6in]{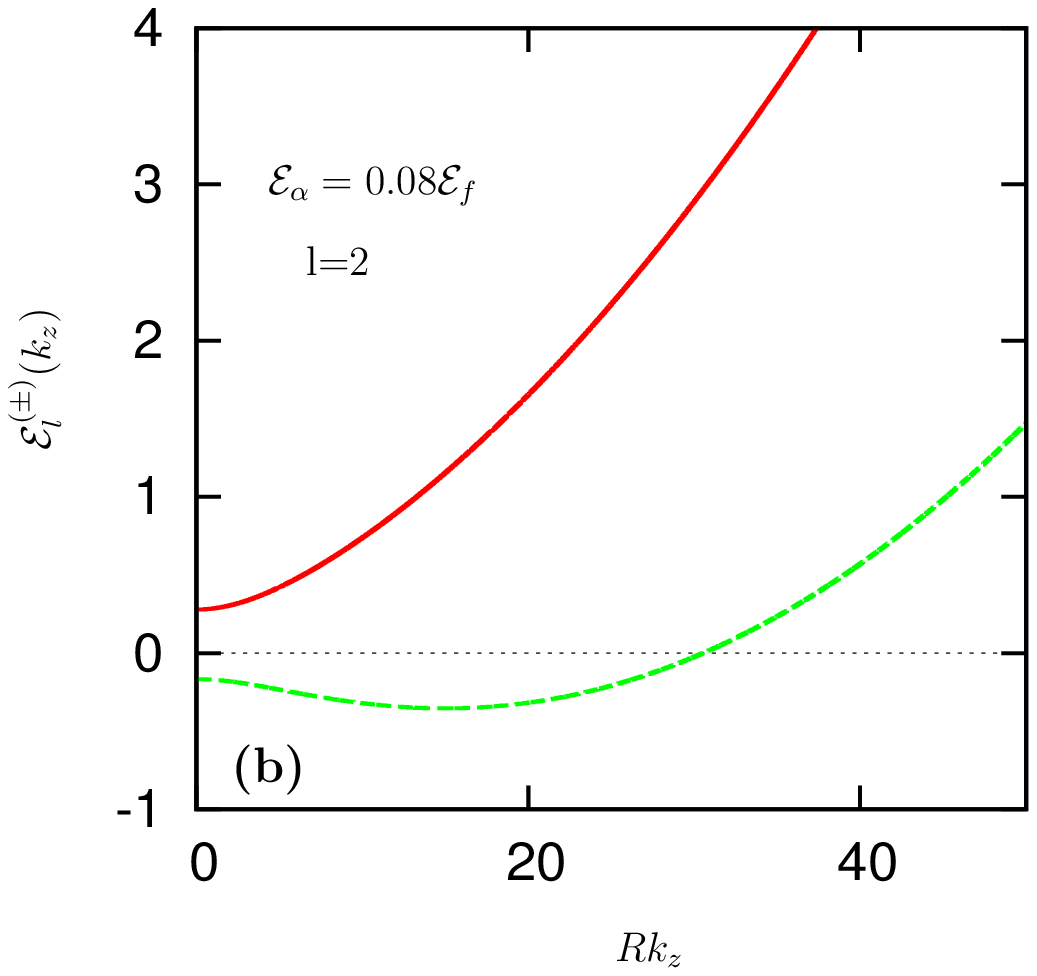}
\caption {(Color online) Energy eigenvalues as a function of $k_z$ in the presence of SOI
${\cal E}_\alpha$. In (a) $l= 0$ and (b) $l=2$ for the angular momentum quantum number. The continuous line is for ``$+$" state while the dashed line is
for ``$-$" state. $R=10$nm is the radius of the nanotube. For $l\ne 0$
((b) $l=2$) there is always an energy gap at $k_z=0$.}
\label{FIG:2}
\end{center}
\end{figure}

The eigenenergies which depend on $k_z$, $l$ and $\alpha_R$ are given by
\begin{eqnarray}
{\cal E}^{s}(k_z,l,\alpha_R)&=&\frac{1}{2}\left[E^{2}_{l+1}(k_z,\alpha_R)
+E^{1}_{l}(k_z,\alpha_R)\right]+\frac{s}{2}\sqrt{\left[E^{2}_{l+1}(k_z,\alpha_R)
-E^{1}_{l}(k_z,\alpha_R) \right]^{2}+4\alpha^{2}_{R}k^{2}_{z} } \ ,
\label{eng1}
\end{eqnarray}
where  $E^{1}_{l}(k_z,\alpha_R)=E^{(0)}(k_z,l)-\frac{\alpha_{R}l}{R}
=E^{(0)}(k_z,l)-l{\cal E}_{\alpha}$
and $E^{2}_{l}(k_z,\alpha_R)=E^{(0)}(k_z,l)+\frac{\alpha_{R}l}{R}
=E^{(0)}(k_z,l)+l{\cal E}_{\alpha}$
with $E^{(0)}(k_z,l)=\frac{\hbar^2k^2_z}{2m^\ast}+\frac{\hbar^2l^2}{2m^\ast R^2}
=\frac{\hbar^2k^2_z}{2m^\ast}+l^2{\cal E}_R$.
Here, $s=\pm$ denotes the two pseudospin orientations. Figures (\ref{FIG:2}a)
and (\ref{FIG:2}b)  depict the dispersion relation given in Eq. (\ref{eng1}) 
exhibiting the split between  the $``+"$ and $``-"$ states. In these
plots, we chose the angular momentum quantum number $l=0$ and $l=2$. At $k_z=0$,
the gap in the spectrum is determined by the SOI as well as the radius of the nanotube.
The corresponding eigenspinors are

\begin{eqnarray}
|\psi_{k_z,L,s}(z,\phi)\rangle=\left(
\begin{array}{ccc}
u^{+}(\phi,z) &\\
u^{-}(\phi,z) &
\end{array}
\right)^{s}_{k_z,L}&=&
\frac{c^{s}_L(k_z,\alpha_R)}{\sqrt{2\pi L_z }}
\left(
\begin{array}{ccc}
1 &\\
\chi^{s}_L(k_z,\alpha_R) &
\end{array}
\right)e^{i(k_zz+L\phi)}\ ,
\label{es1a}
\end{eqnarray}
where \[\chi^{s}_L(k_z,\alpha_R)=e^{i\phi}[E^{1}_{L}(k_z,\alpha_R)-{\cal E}^{s}
(k_z,L,\alpha_R)]/ik_z\alpha_R.  \]
Orthonormality $\langle\psi_{k_z,L^\prime,s^\prime }(z,\phi)|\psi_{k_z,L,s}(z,
\phi)\rangle=\delta_{LL^\prime}\delta_{ss^\prime}$
yields

\begin{eqnarray}
c^{s}_L(k_z,\alpha_R)&=&[1+\chi^{s}_L(k_z,\alpha_R)\chi^{s\ \ast}_L(k_z,
\alpha_R)]^{-1/2}\nonumber\\
&=&k_z\alpha_R[(E^{1}_L(k_z,\alpha_R)-{\cal E}^s(k_z,L,
\alpha_R))^2+\alpha^2_Rk^2_z]^{-1/2}\nonumber\\
\chi^{-}_L(k_z,\alpha_R)&=&\frac{-1}{\chi^{+\ \ast}_L(k_z,\alpha_R)} \ .
\end{eqnarray}

%%%%%%%%%%%%%%%%%%%%%%%%%%%%%%%%%%%%%%%%%%%%%%%%%%

\subsection{Presence of a Barrier}

In the presence of a potential barrier of height $U$ and width $w$, as shown in 
Fig.\ \ref{FIG:1},  the energies and eigenspinors within the barrier
are obtained in a similar fashion as in the absence of a barrier. The effect of the
barrier is to shift energy eigenvalues as follows,

\begin{eqnarray}
{\cal E}^{s}(k^b_z,l,\alpha_R)&\to&{\cal E}^{s}(k^b_z,l,\alpha_R)-U
=\bar{\cal E}^{s}(k^b_z,l,\alpha_R)\nonumber\\
&=&\frac{1}{2}\left[E^{2}_{l+1}(k^b_z,\alpha_R)
+E^{1}_{l}(k^b_z,\alpha_R)\right]
+\frac{s}{2}\sqrt{\left[E^{2}_{l+1}(k^b_z,\alpha_R)
-E^{1}_{l}(k^b_z,\alpha_R) \right]^{2}+4\alpha^{2}_{R}(k^{b}_{z})^2 }.
\label{eng2}
\end{eqnarray}
Outside the barrier ($U=0$), we consider only states with $k_z^2\ge 0$
and we have $e^{\pm ik_z}$ type solutions.
Inside the barrier for small enough barrier heights Eq. (\ref{eng2})implies
$(k^b_z)^2\ge 0$ and we still have $e^{\pm ik^b_zz} $ type solutions.
However as the height of the barrier increases Eq. \ref{eng2} cannot be
satisfied with real $k^b_z$.
In this case, we seek a solution with
$k^b_z\to \pm i\kappa_z$ inside the barrier region giving $e^{\pm \kappa_zz}$
dependence exhibiting decay or growth.

We note that for real $k^b_z$, $\bar{\cal E}^{+}(k^b_z,l,\alpha_R)$ is a
monotonically increasing even function of $k^b_z$, with a minimum at
$k^+_{min}=0$ given by $\bar{\cal E}^{+}(0,l,\alpha_R)=E^2_{l+1}(0,\alpha_R)> 0$,
while $\bar{\cal E}^{-}(k^b_z,l,\alpha_R)$ attains a local maximum at $k^-_{max}=0$
given by $E^1_{l}(0,\alpha_R)=l^2{\cal E}_R-l{\cal E}_{\alpha}$ and can be negative
for high enough ${\cal E}_{\alpha}$. Consequently, we have a finite difference in
energy between the $``+"$ and the $``-"$ states at $k^b_z=0$. For nanotubes 
with large radius ($R\to\infty$), both ${\cal E}_R$ and ${\cal E}_{\alpha}$ go 
to zero and the energy difference between the states vanishes at $k^b_z=0$. For 
real $k^b_z$, $\bar{\cal E}^{-}(k^b_z,l)$ attains a minima at
\[
k^-_{min}R=\pm\left[\left(\frac{{\cal E}_{\alpha}}{{\cal E}_R}\right)^2-
(2l+1)^2\left(1+\frac{{\cal E}_{\alpha}}{{\cal E}_R}\right)\right]^{1/2},
\]
given by $\bar{\cal E}^{-}(k^-_{min},l,\alpha_R)<0$.
However, for high enough potential
$U$, $\bar{\cal E}^{\pm}(k^b_{z},l,\alpha_R)$ in Eq.\  (\ref{eng2}) can have
values less than $\bar{\cal E}^{\pm}(k^{\pm}_{min},l,\alpha_R)$ and $k^b_z$ becomes
imaginary, $k^b_z\to \pm i\kappa_z$  and the state exhibit $e^{\pm \kappa_zz}$
dependence within the barrier. The derivation in this case is the same as the
derivation for the energy eigenvalues and eigenspinors. The eigenenergies are
given by,

\begin{eqnarray}
{\cal E}^{s}(\kappa_z,l,\alpha_R)-U&=&\frac{1}{2}
\left[E^{2}_{l+1}(\kappa_z,\alpha_R)
+E^{1}_{l}(\kappa_z,\alpha_R)\right]+\frac{s}{2}
\sqrt{\left[ E^{2}_{l+1}(\kappa_z,\alpha_R)
-E^{1}_{l}(\kappa_z,\alpha_R) \right]^{2}-4\alpha^{2}_{R}\kappa^{2}_{z} }\ ,
\label{eng3}
\end{eqnarray}
where $E^{1}_{l}(\kappa_z,\alpha_R)=E^{(0)}(\kappa_z,l)-\frac{\alpha_{R}l}{R}$
and $E^{2}_{l}(\kappa_z,\alpha_R)= E^{(0)}(\kappa_z,l)+\frac{\alpha_{R}l}{R}$,
with $E^{(0)}(\kappa_z,l)=\frac{-\hbar^2\kappa^2_z}{2m^\ast}
+\frac{\hbar^2l^2}{2m^*R^2}$. The eigenspinors are similarly modified
with $k_z\to\pm i\kappa$.

%%%%%%%%%%%%%%%%%%%%%%%%%%%%%%%%%%%%%%%%%%%%

\subsection{Limiting case $R\to\infty$}

We show here that the results in the previous section reduce to the familiar
results of 2 DEG with SOI in the limit $R\to\infty$.
For $R\to\infty$, $l/R\to k_{\perp}$ and $l\phi=(l/R)R\phi\to k_{\perp}x_{\perp}$, we obtain for the eigenvalues

\begin{eqnarray}
\lim_{R\to\infty}{\cal E}^{s}(k_z,l)&\to& E^{0}(k_z,l)+s\alpha_R\sqrt{k_{\perp}^2+k_z^2}
\nonumber\\
&=&E^{0}(k_z,l)+s\alpha_R k.
\end{eqnarray}
By making use of

\begin{eqnarray}
\lim_{R\to\infty}\chi^{s}_L(k_z,\alpha_R)&\to&\frac{i(k_{\perp}+sk)}{k_z}\\
\lim_{R\to\infty}c^{s}(k_z,L)&\to&\frac{k_z}{\left[k^2_z
+(k_{\perp}+sk)^2)\right]^{1/2}} \ ,
\end{eqnarray}
 the normalized eigenspinors become

\begin{eqnarray}
|\psi^{+}(z,x_{\perp})\rangle=\frac{k_z}{\sqrt{k_z^2+(k_{\perp}+k)^2}}
\left(
\begin{array}{ccc}
1 &\\
\frac{i(k_{\perp}+k)}{k_z} &
\end{array}\right)\frac{e^{i(k_zz+k_{\perp}x_{\perp})}}{\sqrt{A}}
\nonumber\\
|\psi^{-}(z,x_{\perp})\rangle=\frac{k_z}{\sqrt{k_z^2+(k_{\perp}-k)^2}}
\left(
\begin{array}{ccc}
1\\
\frac{i(k_{\perp}-k)}{k_z} &
\end{array}\right)\frac{e^{i(k_zz+k_{\perp}x_{\perp})}}{\sqrt{A}} \ ,
\end{eqnarray}
where $A=2\pi RL_z$, where $L_z$ is a normalization length.
At first glance,  these expressions look different from
the 2DEG expressions for the eigenspinors. The reason for this is that the
geometry used here is different (rotated). It can be shown that if we start with
a different geometry for the 2DEG confinement we obtain the above result.
Here, in the limit $R\to\infty$, we have a confinement in the $y-z$ plane and
the SO Hamiltonian for the 2DEG is written as

\begin{eqnarray}
H_{SO}&=&\frac{\hbar}{(2m^\ast c)^2}{\bf\nabla} V_x
\cdot(\mbox{\boldmath$\sigma$}\times{\bf p})_x
\nonumber\\
&=&i\alpha\left(\sigma_3\frac{\partial}{\partial y}
-\sigma_2\frac{\partial}{\partial z}\right) \ .
\end{eqnarray}
This leads  to the following eigenvalue problem

\begin{equation}
\left(\matrix{E^{(0)}-\alpha k_y& -ik_z\alpha\cr
ik_z\alpha & E^{(0)}+\alpha k_y\cr}
\right)\left(\matrix{c^{+}\cr
c^{-}\cr}
\right)={\cal E}\left(\matrix{c^{+}\cr
c^{-}\cr}
\right)\ .
\end{equation}
Here, $E^{(0)}=\frac{\hbar^2}{2m^\ast }(k_y^2+k_z^2)=\frac{\hbar^2k^2}{2m^\ast}$, and
the eigenvalue equation is given by

\begin{eqnarray}
&&\left((E^{(0)}-{\cal E})-\alpha k_y\right)
\left((E^{(0)}-{\cal E})+\alpha k_y\right)-\alpha^2 k_z^2=0
\nonumber\\
&&({\cal E}-E^{(0)})^2-\alpha^2(k_y^2+k_z^2)=0
\nonumber\\
&&{\cal E}^{s}=E^{(0)}+s\alpha k \ .
\end{eqnarray}
The eigenspinors are given by

\begin{eqnarray}
{\cal E}^{+}:\;\;\;\;\;&&(E^{(0)}-\alpha k_y)c^{+}-i\alpha k_zc^{-}={\cal E}^{+}c^{+}
\nonumber\\
&&c^{-}=({\cal E}^{+}-E^{(0)}+\alpha k_y)\frac{c^{+}}{-i\alpha k_z}=\frac{i(k_y+k)}{k_z}c^{+}
\nonumber\\
{\cal E}^{-}:\;\;\;\;\;&&(E^{(0)}+\alpha k_y)c^{-}+i\alpha k_zc^{+}={\cal E}^{-}c^{-}
\nonumber\\
& &c^{+}=({\cal E}^{-}-E^{(0)}-\alpha k_y)\frac{c^{-}}{i\alpha k_z}=\frac{i(k_y-k)}{k_z}c^{-} \ .
\end{eqnarray}
These lead to the eigenspinors obtained above in the limit $R\to\infty$.
The difference in appearance is simply to the choice of the direction of
confinement and all the eigenspinors are equivalent.

%%%%%%%%%%%%%%%%%%%%%%%%%%%%%%%%%%%%%%%%%%%%%%%%%%%%%%%%%%

\section{Tunneling through a Potential Barrier}
\label{sec3}

We consider a potential barrier localized on the nanotube. The
potential has height $U$ and width $w$. We calculate the transition
probability of an electron propagating in the $+z$-direction.
We have three regions of interest. In region I, $z<0$, we
consider a specific incoming state (a linear combination
of the up and down-spin eigenstates) with wave vector $k^{i}_z$ and
angular momentum quantum $L^{i}$ as well as the  reflected state with wave
vector $-k^{i}_{z}$, with two possible spin states. Therefore, in
region I, before the barrier, the wavefunction is

\begin{eqnarray}
|\psi^{1}_{k^{i}_z,L^{i}}\rangle&=&a_{+}|k^{i}_z,L^{i},+\rangle+a_{-}
|k^{i}_z,L^{i},-\rangle+\sum_{ss^\prime=\pm}
r_{ss^\prime}|-k^{i}_z,L^{r},ss^\prime\rangle.
\nonumber\\
1&=&|a_+|^2+|a_-|^2 \ .
\end{eqnarray}
That is, in region I, we have an incoming superposition of $\pm$ eigenspinors
which are normalized and a reflected superposition of $\pm$ states.
In region III, after the barrier we have only a transmitted state given by

\begin{equation}
|\psi^{3}_{k^{t}_z,L^{t},s}\rangle=\sum_{ss^\prime=\pm}t_{ss^\prime}|
k^{i}_z,L^{t},ss^\prime\rangle \ .
\end{equation}
In region II, inside the barrier, the form of the wavefunction is the
same as in Eq.\ (\ref{es1a}) with a different $k_z$  due to the
presence of the potential barrier and we have

\begin{equation}
|\psi^{2}_{k^{b}_z,L^{b},s}\rangle=\sum_{rs^\prime=\pm}b_{rs^\prime}|
rk^{b}_z,L^{b},rs^\prime \rangle \ .
\end{equation}
Here, $r_{ss^\prime}$, $t_{ss^\prime}$ and $b_{rr^\prime}$ are the matrix
elements  associated with the scattering process corresponding to reflection,
transmission and barrier states. They are determined using the appropriate 
boundary conditions.

Let us consider the case when the incoming electron is
given by a superposition states of Eq. (\ref{es1a}).  Continuity
of $\psi$ at $z=0$ gives

\begin{eqnarray}
&&\sum_{s=\pm}a_sc^{s}_{L^i}(k^i_z)\left(
\begin{array}{ccc}
1 &\\
\chi^{s}_{L^i}(k^i_z) &
\end{array}
\right)e^{iL^i\phi}+\sum_{ss^\prime=\pm}r_{-ss^\prime}c^{ss^\prime}_{L^r}(k^i_z)\left(
\begin{array}{ccc}
1 &\\
\chi^{ss^\prime}_{L^r}(-k^i_z) &
\end{array}
\right)e^{iL^r\phi}\\
&=&\sum_{ss^\prime=\pm}b_{+ss^\prime}c^{ss^\prime}_{L^b}(k^b_z)\left(
\begin{array}{ccc}
1 &\nonumber\\
\chi^{ss^\prime}_{L^b}(k^b_z) &
\end{array}
\right)e^{iL^b\phi}
+\sum_{ss^\prime=\pm}b_{-ss^\prime}c^{ss^\prime}_{L^b}(k^b_z)\left(
\begin{array}{ccc}
1 &\\
\chi^{ss^\prime}_{L^b}(-k^b_z) &
\end{array}
\right)e^{iL^b\phi} \ .
\end{eqnarray}
For the above to be true for all values of $\phi$, we must have $L^b=L^r=L^i$,
which is due to conservation of angular momentum along the $z$-axis,
which is case for a potential that is
independent of $\phi$. Continuity of $\psi^\prime$ at $z=0$ yields

\begin{eqnarray}
&&k^i_z\sum_{s=\pm}a_sc^{s}_{L^i}(k^i_z)\left(
\begin{array}{ccc}
1 &\\
\chi^{s}_{L^i}(k^i_z) &
\end{array}
\right)-k^i_z\sum_{ss^\prime=\pm}r_{-ss^\prime}c^{ss^\prime}_{L^i}(k^i_z)\left(
\begin{array}{ccc}
1 &\\
\chi^{ss^\prime}_{L^i}(-k^i_z) &
\end{array}
\right)\\
&=&k^b_z\sum_{ss^\prime=\pm}b_{+ss^\prime}c^{ss^\prime}_{L^i}(k^b_z)\left(
\begin{array}{ccc}
1 &\nonumber\\
\chi^{ss^\prime}_{L^i}(k^b_z) &
\end{array}
\right)
-k^b_z\sum_{ss^\prime=\pm}b_{-ss^\prime}c^{ss^\prime}_{L^i}(k^b_z)\left(
\begin{array}{ccc}
1 &\\
\chi^{s^\prime}_{L^i}(-k^b_z) &
\end{array}
\right) \ .
\end{eqnarray}
Continuity of $\psi$ at $z=w$ gives us

\begin{eqnarray}
&&\sum_{ss^\prime=\pm}b_{+ss^\prime}c^{ss^\prime}_{L^i}(k^b_z)\left(
\begin{array}{ccc}
1 &\nonumber\\
\chi^{ss^\prime}_{L^i}(k^b_z) &
\end{array}
\right)e^{ik^b_zw}
+\sum_{ss^\prime=\pm}b_{-ss^\prime}c^{ss^\prime}_{L^i}(k^b_z)\left(
\begin{array}{ccc}
1 &\\
\chi^{ss^\prime}_{L^i}(-k^b_z) &
\end{array}
\right)e^{-ik^b_zw}
\nonumber\\
&=&\sum_{ss^\prime=\pm}t_{+ss^\prime}c^{ss^\prime}_{L^i}(k^i_z)\left(
\begin{array}{ccc}
1 &\nonumber\\
\chi^{ss^\prime}_{L^i}(k^i_z) &
\end{array}
\right)e^{ik^i_zw} \ .
\end{eqnarray}

Continuity of $\psi^\prime$ at $z=w$ leads to

\begin{eqnarray}
&&k^b_z\sum_{ss^\prime=\pm}b_{+ss^\prime}c^{ss^\prime}_{L^i}(k^b_z)\left(
\begin{array}{ccc}
1 &\nonumber\\
\chi^{ss^\prime}_{L^i}(k^b_z) &
\end{array}
\right)e^{ik^b_zw}
-k^b_z\sum_{ss^\prime=\pm}b_{-ss^\prime}c^{ss^\prime}_{L^i}(k^b_z)\left(
\begin{array}{ccc}
1 &\\
\chi^{ss^\prime}_{L^i}(-k^b_z) &
\end{array}
\right)e^{-ik^b_zw}
\nonumber\\
&=&k^i_z\sum_{ss^\prime=\pm}t_{+ss^\prime}c^{ss^\prime}_{L^i}(k^i_z)\left(
\begin{array}{ccc}
1 &\nonumber\\
\chi^{ss^\prime}_{L^i}(k^i_z) &
\end{array}
\right)e^{ik^i_zw} \ .
\end{eqnarray}
We solve these  coupled equations for the reflection
and transmission matrices. Here, depending on the energy of
the incoming electrons and the height  of the potential $U$,
$k^b_i$ can be real or imaginary. The above boundary conditions
yield eight equations for the eight coefficients.

%%%%%%%%%%%%%%%%%%%%%%%%%%%%%%%%%%%%%%%%%%%%%%%%%%%%%%%%%%%%%%%%%%

\subsection{Transmission and Reflection Amplitudes}

We  are interested in obtaining the  transmission and reflection
probabilities which will be used below in our numerical calculations.
The solutions for the probability amplitudes for all barrier height are
given by,

\begin{equation}
t_{+\pm} = - \frac {4a_{\pm}e^{iw \left( k_z^b - k_z^i \right)} k_z^b k_z^i }
{{e^{2iwk_z^b } \left( {k_z^b  - k_z^i } \right)^2  - \left( {k_z^b  + k_z^i } \right)^2 }}
\end{equation}

\begin{equation}
r_{ -\pm }  = \frac{{\left[ {\left( {k_z^b } \right)^2
  - \left( {k_z^i } \right)^2 } \right]\Big\{2a_{\mp}c^{\mp}(k^i_z)\chi^-(k^i_z)+a_{\pm}c^{\pm}(c^i_z)\left[ {\chi ^ -  \left( {k_z^i }
    \right) + \chi ^ +  \left( {k_z^i } \right)} \right]\Big\}\sin \left( {wk_z^b } \right)}}
{c^{\pm}(k^i_z){\left[ {\chi ^ -  \left( {k_z^i } \right) - \chi ^ +  \left( {k_z^i } \right)}
    \right]\left\{ {2ik_z^b k_z^i \cos \left( {wk_z^b } \right)
  + \left[ {\left( {k_z^i } \right)^2  + \left( {k_z^b } \right)^2 } \right]\sin
    \left( {wk_z^b } \right)} \right\}}} \,
\end{equation}
where depending on the barrier height $k_z^b\to i\kappa$.
The transmission and reflection probabilities for real $k^b_z$ are given by

 \begin{eqnarray}
|t_{+\pm}|^2&=&\frac{16|a_{\pm}|^2(k^b_z)^2(k^i_z)^2}{(k^i_z+k^b_z)^4+(k^i_z-k^b_z)^4\
-2(k^b_z-k^i_z)^2(k^b_z+k^i_z)^2\cos(2wk^b_z)}
\nonumber\\
&=&\frac{16|a_{\pm}|^2(k^b_z)^2(k^i_z)^2}{2(k^i_z)^4+2(k^b_z)^4+12(k^i_z)^2(k^b_z)^2
-2(k^b_z-k^i_z)^2(k^b_z+k^i_z)^2\cos(2wk^b_z)}
\nonumber\\
|r_{-\pm}|^2&=&
\frac{4\big(|a_+|^2+|a_-|^2\big)\left [(k^i_z)^2-(k^b_z)^2\right]^2\sin^2(wk^b_z)}
{\Big\{(k^i_z+k^b_z)^4+(k^i_z-k^b_z)^4
-2(k^b_z-k^i_z)^2(k^b_z+k^i_z)^2\cos(2wk^b_z)\Big\}}
\label{prob1}
\end{eqnarray}
  for all values of   $a_+$ and $a_-$, and they satisfy the usual probability
  conservation law
\begin{equation}
|t_{++}|^2+ |t_{+-}|^2+|r_{-+}|^2+|r_{--}|^2=1.
\label{consrv}
\end{equation}
The implication of Eq. (\ref{prob1}) is that we have no
interference term arising from spin flip which is  consistent with a
non-spin-dependent scattering potential assumed in the problem.

For $k^b_z\to\pm i\kappa$, the expressions for the probability amplitudes and
probabilities take the following form

\begin{eqnarray}
t_{+\pm}&=&-\frac{4ia_{\pm}e^{(-\kappa^b_z-ik^i_z)w}\kappa^b_zk^i_z}
{e^{-2\kappa^b_zw}\left(i\kappa^b_z-k^i_z \right)^2
-\left(i\kappa^b_z+k^i_z\right)^2}
\nonumber\\
r_{-\pm}&=&-\frac{\left[(\kappa^b_z)^2+(k^i_z)^2 \right]\left\{
2a_{\mp}c^{\mp}(k^i_z)\chi^-(k^i_z)+a_{\pm}c^{\pm}(k^i_z)
\left[\chi^-(k^i_z)+\chi^+(k^i_z) \right]\right\}\sinh(w\kappa^b_z)}
{c^{\pm}(k^i_z)\left[\chi^-(k^i_z)-\chi^+(k^i_z)\right]\left\{
-2k^i_z\kappa^b_z\cosh(w\kappa^b_z)
+i\left[(k^i_z)^2+(\kappa^b_z)^2 \right]\sinh\left(w\kappa^b_z\right)\right\}},
\end{eqnarray}
and they also satisfy the conservation of probability relation given by Eq. 
(\ref{consrv}).

%%%%%%%%%%%%%%%%%%%%%%%%%%%%%%%%%%%%%%%%%%%%%%%%%%%%%%%%%%%%%%%%%%%%%%%%%%%%%%%%%%%%%%%%%%

\section{Influence of Disorder and Interface Roughness on Tunneling}
\label{newsec4a}

 We now investigate a  model which  determines how disorder in the  potential barrier
  affects the tunneling. We consider a simple  model which assumes  the existence of
interface roughness and shows that the contribution to the tunneling current depends on the
relative strength  between the spin-orbit  coupling on the nanotube and the disorder at  the interface.
 We will study how localization  may  dramatically affect the spin polarization current
 which we calculated in Figs. \ref{FIG:3} and \ref{FIG:4}.
The strong sensitivity of the tunneling spin polarization  to the interface
 structure allows for the possible role which interface roughness might play in  device applications.

In the preceding formalism, we did not include any  considerations of the way
in which  the SO coupling strength in this system affects the transmission
and reflection coefficients  in the presence  of   disorder.  Of course, disorder
 will  give rise to  interface states which would in turn affect the tunneling. These
 realistic considerations are difficult to include beyond the standard
 procedure of diagonalizing the model Hamiltonian.  We include  two effects; one
is  the breakdown of momentum conservation arising from impurity scattering
which is included as a finite lifetime of the electron states,  and
the other is when the impurity scattering is simulated by a Maxwellian
distribution involving a thermal parameter. Both
of these corrections are likely to transfer energy from one state with large angular momentum
 to another state with smaller angular momentum. For the non-conservation of momentum,
we employ a phenomenological approach along the lines  of Marmorkos, Wang and
Das Sarma\cite{das-sarma-1,das-sarma-2} who calculated the polarization function beyond the
random-phase approximation (RPA) to include the effects due to disorder.
They introduced a broadening function which
couples the polarized spectrum for wave number $q$ to that at momentum $q^\prime$.

If the tunneling is assisted by impurities in the barrier, then this is a
second-order process,  involving a  nanotube-to-impurity-to-nanotube  gap
state.\cite{dan1,dan2} In the case when  the tunneling is {\em not} assisted
by impurities in the barrier, it is just a broadening of the nanotube-to-nanotube
tunneling probability due to momentum dissipation. We may include disorder into our
tunneling probabilities through

\begin{equation}
t_{s,s^\prime}^{ \rm disorder}(l)= \frac{ \sum_{l^\prime=-\infty}^\infty e^{-(l-l^\prime)^2/ \Gamma^2}
 t_{s,s^\prime}(l^\prime) }{\sum_{l^\prime=-\infty}^\infty  e^{-(l-l^\prime)^2/ \Gamma^2} } \ ,
\label{broaden}
\end{equation}
where $\Gamma$ is a dimensionless  phenomenological  ``temperature"
parameter representing the degree to which there is a    breakdown in
momentum conservation. As expected, when $\Gamma\to 0$, we recover the original
tunneling probability. Below, in Section \ref{sec4},  we show our numerical  results
  for this effect by  substituting our derived results  for
$t_{s,s^\prime}(l)$  into Eq.\ (\ref{broaden}). We find that finite $\Gamma$
does reduce the transmission probability.

Alternatively, we may simulate the disorder  by

\begin{eqnarray}
t_{s,s^\prime}^{ \rm disorder}(l)&=&\frac{1}{{\cal N}(\gamma)}\frac{1}{\gamma}\sum_{l^\prime=-\infty}^\infty
\int_0^{2\pi} \frac{d\varphi}{2\pi} \int_0^{2\pi} \frac{d\varphi^\prime}{2\pi} \
\frac{e^{i(l-l^\prime)(\varphi-\varphi^\prime)}}{(\varphi-\varphi^\prime)^2+\frac{1}{\gamma^2}}
t_{s,s^\prime}(l^\prime) \ ,
\label{broaden2}
\end{eqnarray}
where ${\cal N}(\gamma)=\sum_{l^\prime=-\infty}^\infty
\int_0^{2\pi} \frac{d\varphi}{2\pi} \int_0^{2\pi} \frac{d\varphi^\prime}{2\pi} \
\frac{e^{i(l-l^\prime)(\varphi-\varphi^\prime)}}
{(\varphi-\varphi^\prime)^2+\frac{1}{\gamma^2}}$ is a normalization factor
and in this case $\gamma$ plays the role of inverse lifetime.

There have been some calculations on the role played by defects on enhancing
electron tunneling through barriers.\cite{dan1,dan2}   These authors
calculated the capture probability of free electrons by defects located in
the barrier and the subsequent emission probabilities of the captured electrons
by thermal emission or phonon-assisted tunneling. However, the defect was assumed
fixed in position so that the effect of disorder was not included  in those calculations.

\section{Numerical Results and Discussion}
\label{sec4}

\begin{figure}[p]
\begin{center}
\includegraphics[width=3.6in]{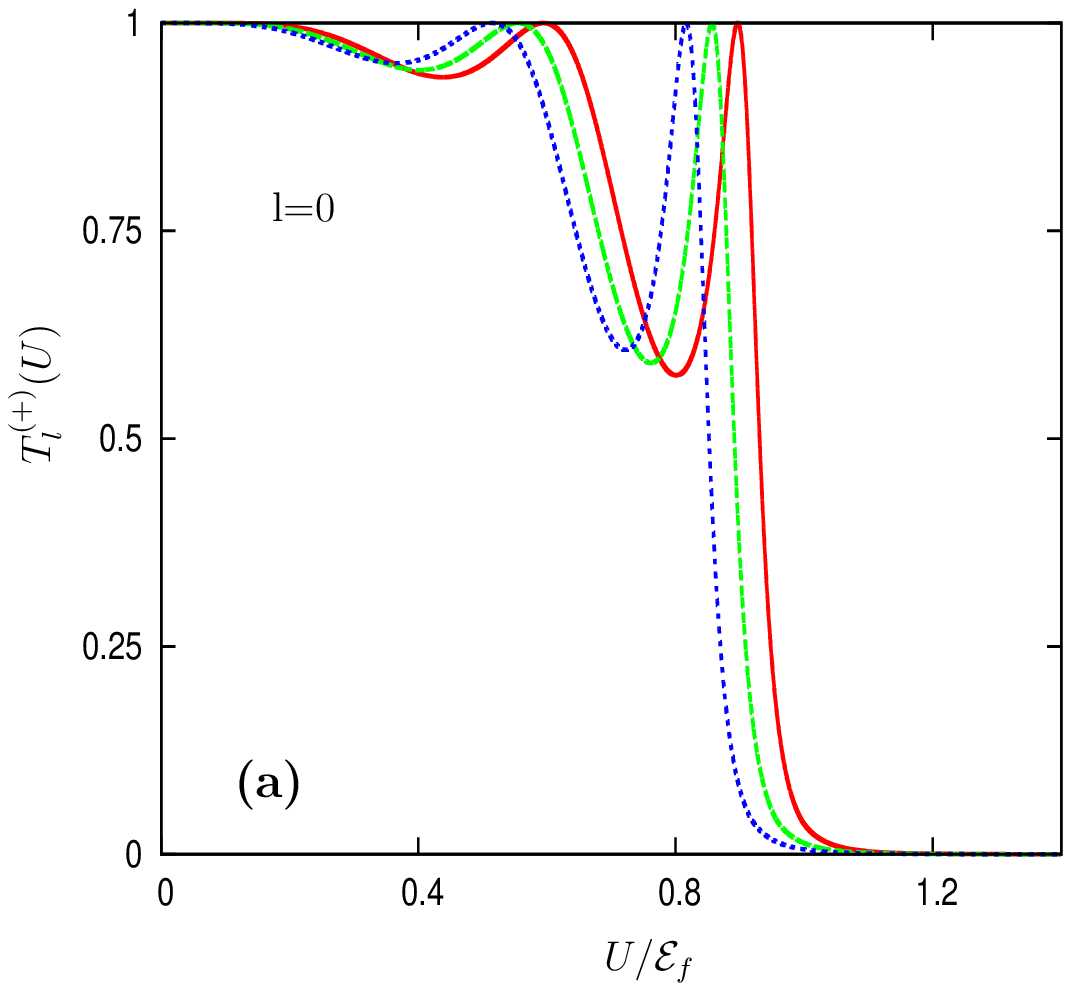}
\includegraphics[width=3.6in]{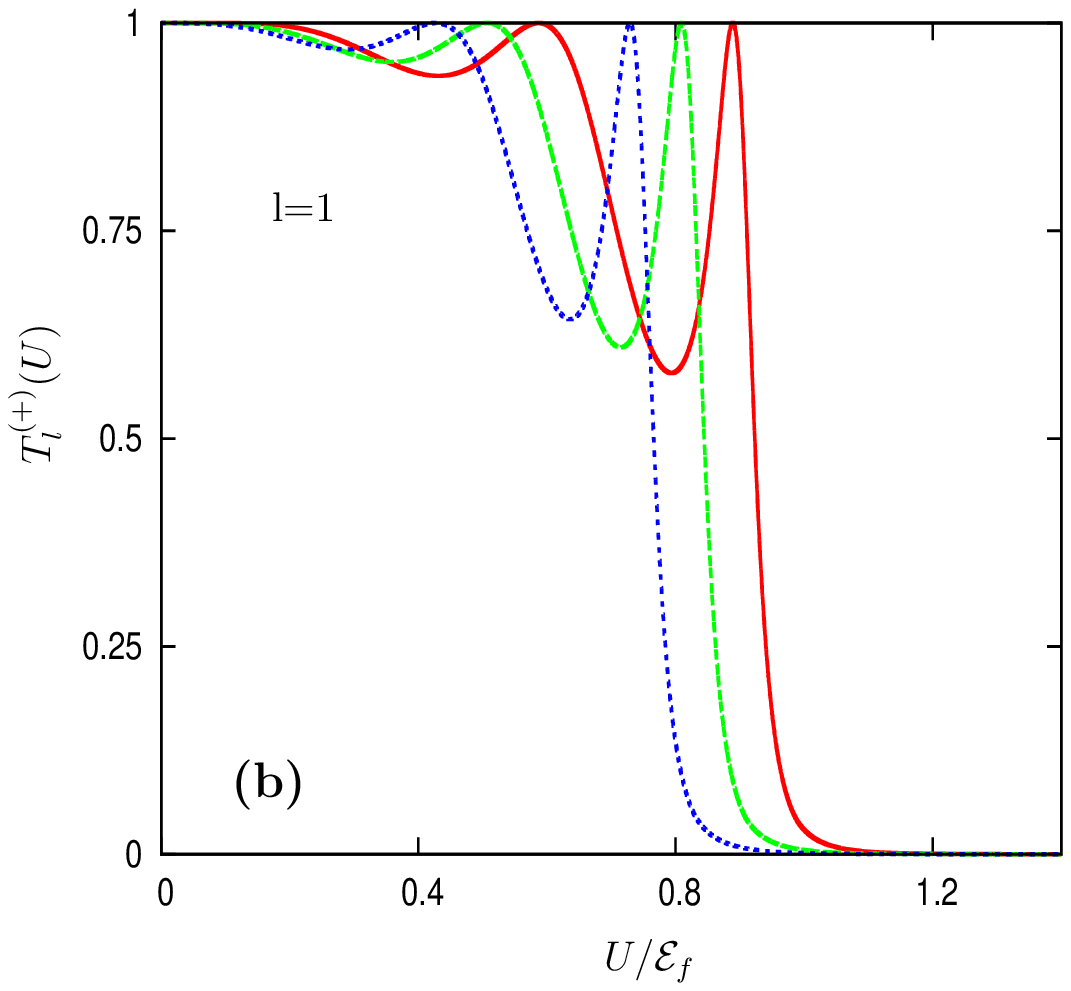}
\caption {( (Color online) Transmission probabilities for incident electrons in
the the ``$+$" state  as functions of the  potential height $U$ for three values of SOI energy
${\cal E}_{\alpha}$ for (a) $l=0$ and (b) $l=1$.  Here,
${\cal E}_{\alpha}=0$ (solid line),
$0.04{\cal E}_f$ (dashed line) and $0.08{\cal E}_f$ (dotted line) where
the incident energy $E^i={\cal E}_f$. }
\label{FIG:3}
\end{center}
\end{figure}

\begin{figure}[p]
\begin{center}
\includegraphics[width=3.6in]{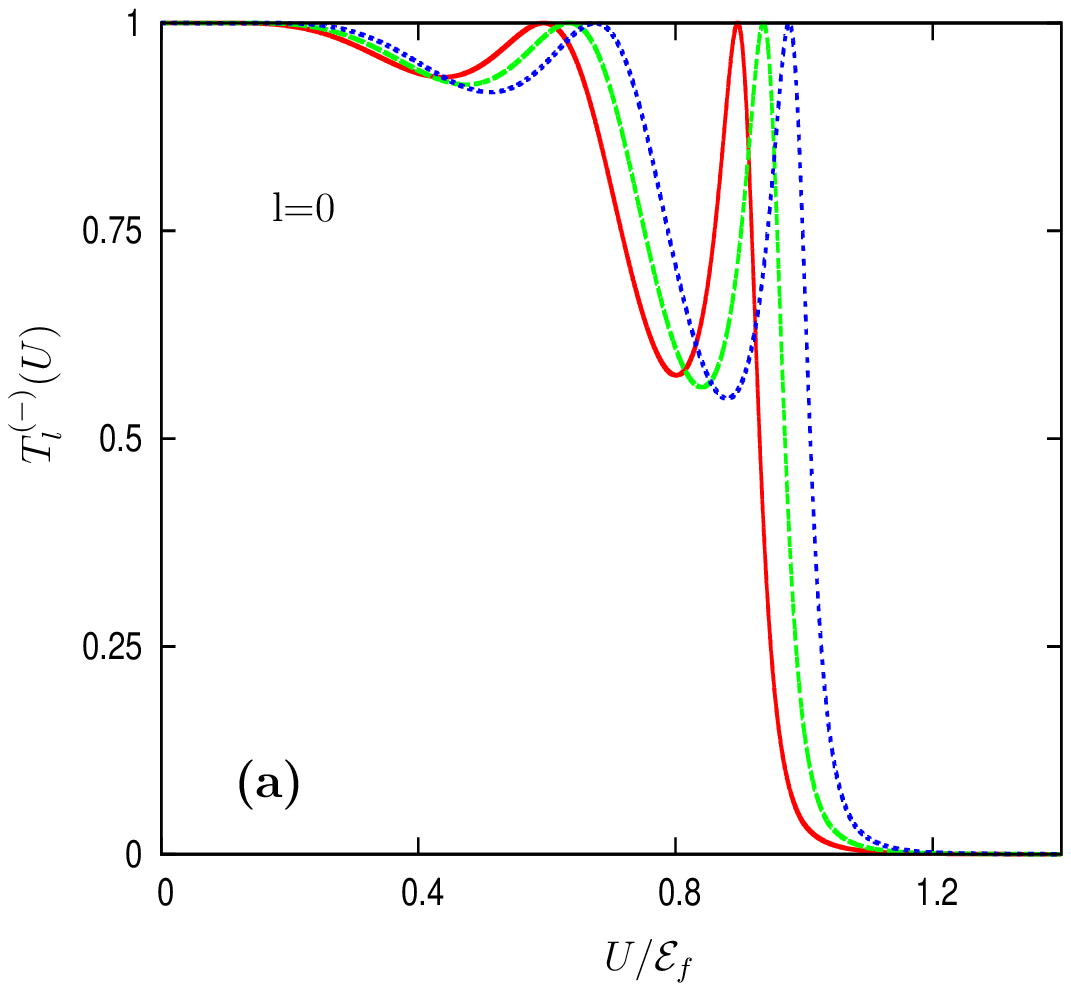}
\includegraphics[width=3.6in]{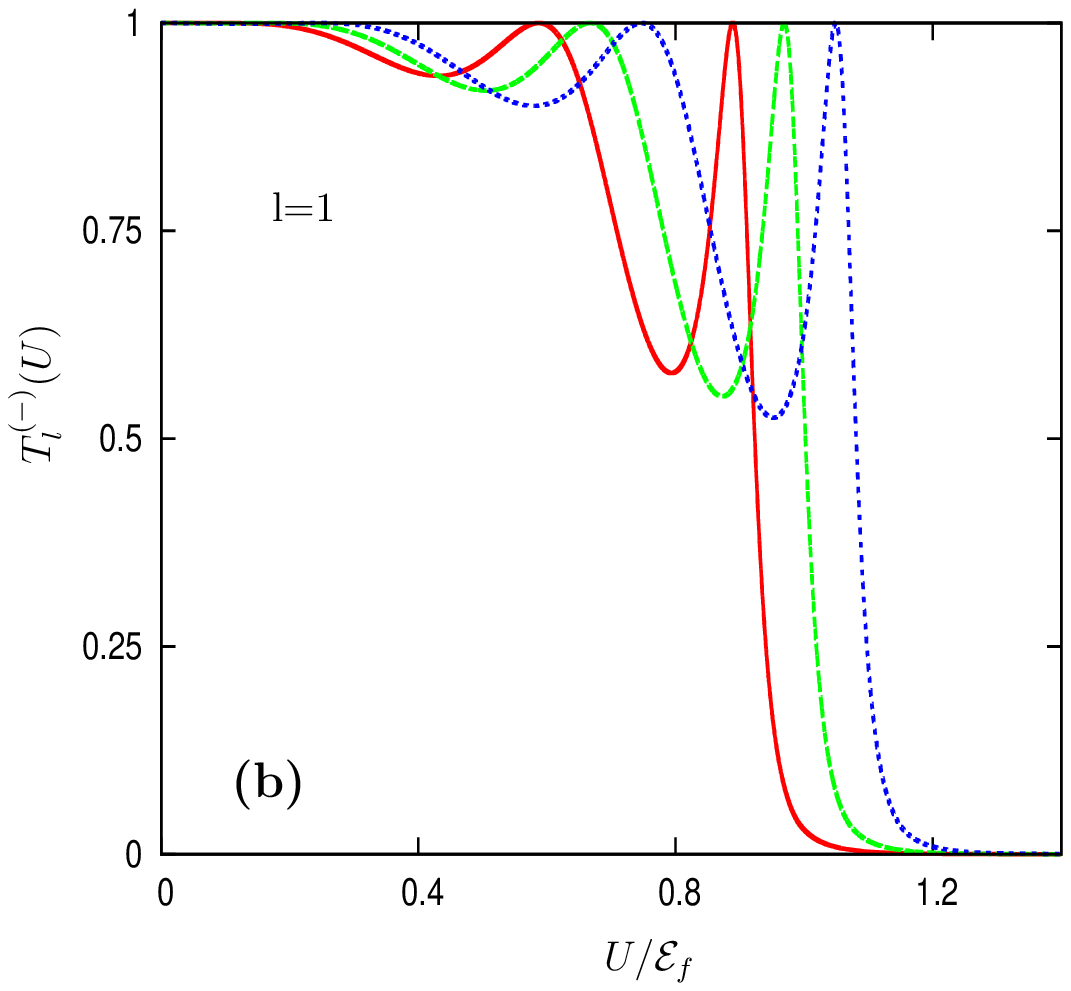}
\caption {(Color online) Transmission probabilities
for electrons in the  ``$-$" state as functions of the   height $U$ of the
potential barrier for three values of SOI energy
${\cal E}_{\alpha}$, In (a) $l=0$ and (b) $l=1$ and
we chose  ${\cal E}_{\alpha}=0$ (solid line),
$0.04{\cal E}_f$ (dashed line) and $0.08{\cal E}_f$ (dotted line).}
\label{FIG:4}
\end{center}
\end{figure}

\begin{figure}[p]
\begin{center}
\includegraphics[width=3.6in]{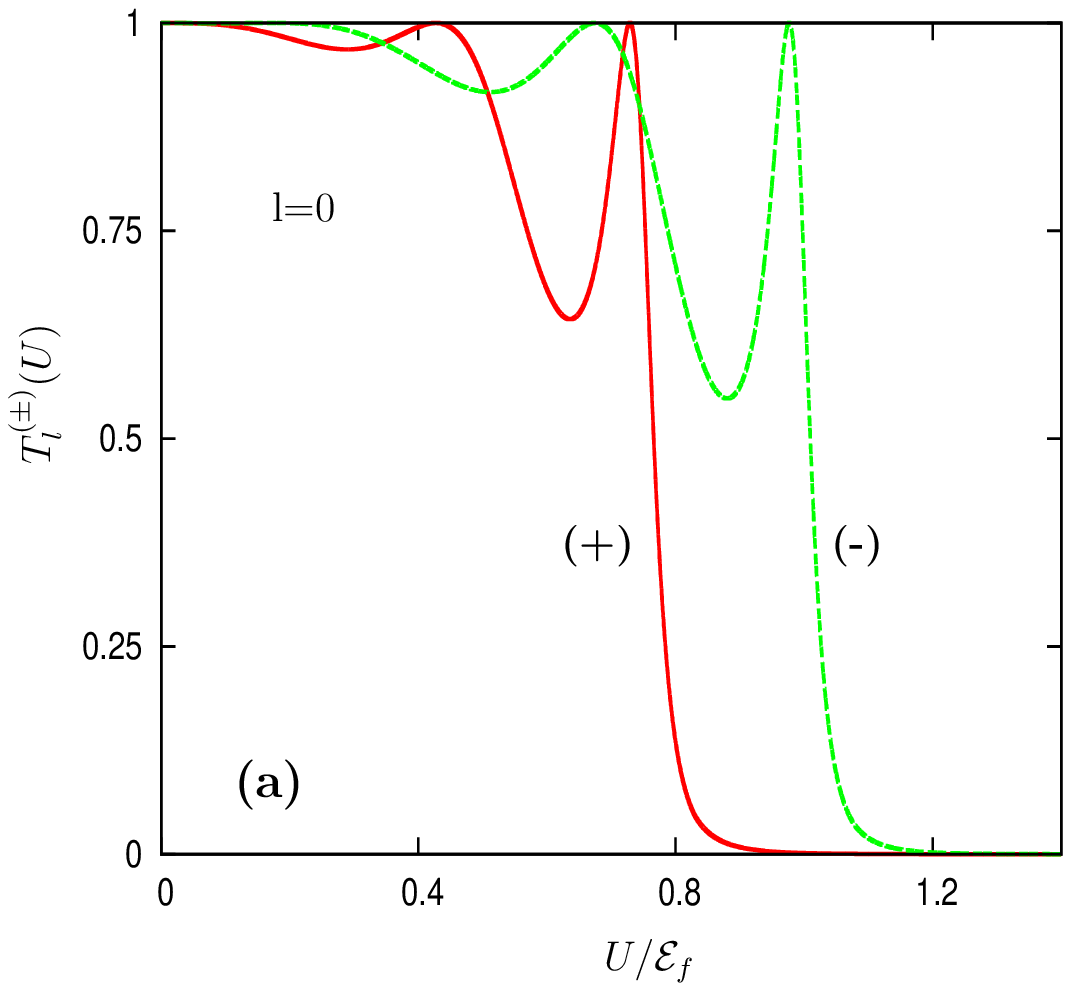}
\includegraphics[width=3.6in]{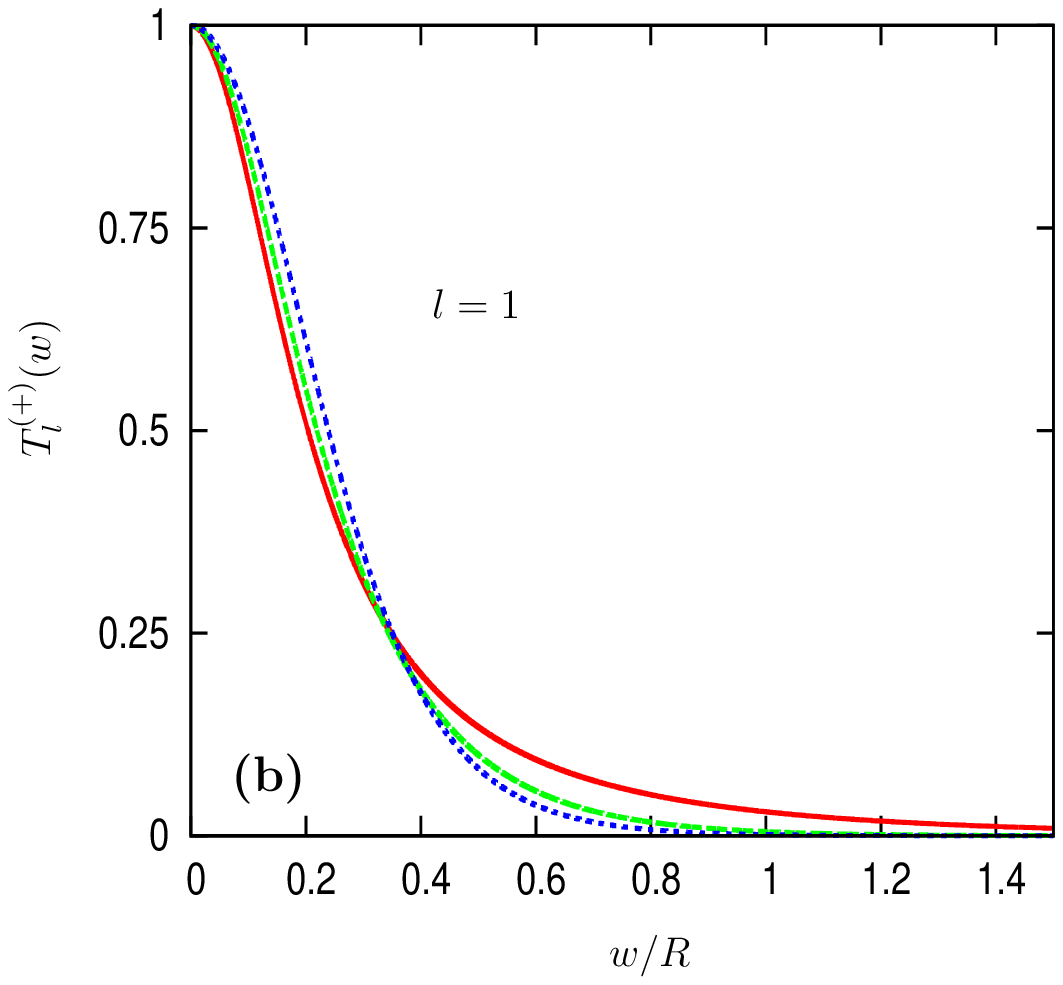}
\caption {(Color online)  (a) Comparison of the transmission probabilities
 between ``$+$" and ``$-$" states for ${\cal E}_{\alpha}=0.08{\cal E}_f$
 and angular momentum quantum number $l=0$.  (b) The dependence of the transmission
 probability on the width of the   barrier $w$ in units of $R$ the radius of the
 nanotube ($R=10$nm) for   ``$+$" state. We chose ${\cal E}_{\alpha}=0
 $ (solid line), $0.04{\cal E}_f$ (dashed line) and $0.08{\cal E}_f$ (dotted  line). }
\label{FIG:5}
\end{center}
\end{figure}

In order to analyze the dependence of the tunneling probability
on the potential barrier height, width and incident energy, we examine the energy
eigenvalue dispersion relation. For chosen  values of incident energy and angular
momentum, we have a range of allowed wave vectors consistent with conservation of
energy.  We invert the energy relationships given in Eqs. (\ref{eng2}) and (\ref{eng3})
($k^b_z\to i\kappa$)

\begin{eqnarray}
\bar{\cal E}^{s}(k^b_z,l,\alpha_R)&=&\frac{1}{2}\left[E^{2}_{l+1}(k^b_z,\alpha_R)
+E^{1}_{l}(k^b_z,\alpha_R)\right]
+\frac{s}{2}\sqrt{\left[E^{2}_{l+1}(k^b_z,\alpha_R)
-E^{1}_{l}(k^b_z,\alpha_R) \right]^{2}+4\alpha^{2}_{R}(k^{b}_{z})^2 }.
\end{eqnarray}
and obtain an expression for $k^b_z(k^i_b,{\cal E}_{\alpha},l,U)$ as a function of
the incident energy $E^i={\cal E}_f$ via $k^i_z$, SOI energy ${\cal E}_{\alpha}$, angular momentum quantum
number $l$ and barrier height $U$.
We plotted in Figs.\  \ref{FIG:3} and \ref{FIG:4} the transmission probabilities as
functions of the potential barrier height and specific values of incident
wave vector $k^i_z$, angular momentum quantum number $l=0,1$ and SOI energy 
${\cal E}_{\alpha}$, for the $``+"$ and $``-"$ states, respectively. 
We note that the
transmission probability exhibits oscillatory behavior where, for certain barrier
height, we have perfect transparency. This is consistent with what is expected of
the result for scattering from a potential barrier in the absence of disorder.
However, the barrier height where
perfect transmission occurs  depends on the values of ${\cal E}_{\alpha}$
and $l$ as well as whether the state is $``+"$ or $``-"$.
This means that we may use the SOI as a filter for obtaining unimpeded transport
through a specified potential barrier height. Furthermore, due to the energy
splitting between $\bar{\cal E}^{\pm}(k^b_z,l,\alpha_R)$ energies,
scattering with SOI can filter the $``+"$ and  $``-"$ states
for clean potential barriers.
When the potential barrier height $U$ is increased so that $\bar{\cal E}^{\pm}
(k^{b}_{z},l,\alpha_R)<\bar{\cal E}^{\pm}(k^{\pm}_{min},l,\alpha_R)$, the
transmission probability shown in Figs.\  \ref{FIG:3} and
\ref{FIG:4} decreases monotonically, while showing a dependence on the SOI.
 Interestingly in Figs. (\ref{FIG:3}a) and (\ref{FIG:3}b), it is shown that
 for the $``+"$ state as the SOI
energy  ${\cal E}_{\alpha}$ is increased, the transmission probability is suppressed
in the sense that it starts to decrease monotonically for smaller barrier heights.
In Figs.\ \ref{FIG:4}(a) and \ref{FIG:4}(b), we plot the transmission probability as a
function of barrier height $U$ for the $``-"$ state for the same values of SOI
energy  ${\cal E}_{\alpha}$ and $l=0,1$. In this case, as the SOI
energy is increased, the transmission probability is increased so that it will start to
decrease monotonically for higher values of the potential height, in effect
allowing for filtering between the two states. Comparison of this effect due to
the SOI energy  ${\cal E}_{\alpha}$ on the transmission probability for the
$``+"$ and $``-"$ states is shown explicitly in Fig. \ref{FIG:5}(a) for chosen
$l$ and ${\cal E}_\alpha$. In Fig. (\ref{FIG:5}b) the dependence of the
transmission   probability as
a function of the barrier width $w$ is given, showing the usual rapid decrease as the
width increases.

\par
We calculated  the transmission probability as a a function of the potential height
$U$ for different values of $l$ with the same ${\cal E}_{\alpha}$ as in Figs.
\ref{FIG:3} - \ref{FIG:5}. The results obtained are qualitatively similar to the
case when the angular momentum quantum number $l=1$ was used. The difference is that
unimpeded transmission occurs at a higher potential height $U$ for $l=0$ compared
with $l=1$. As the value of $l$ increases perfect transmission
occurs at lower potential heights.  This is expected since the energy of
an incoming electron  is divided between the linear and angular motion, yielding
less energy along the axis to the impinging electron, for chosen total energy,
when the angular momentum is increased.

\par
\begin{figure}[p]
\begin{center}
\includegraphics[width=3.6in]{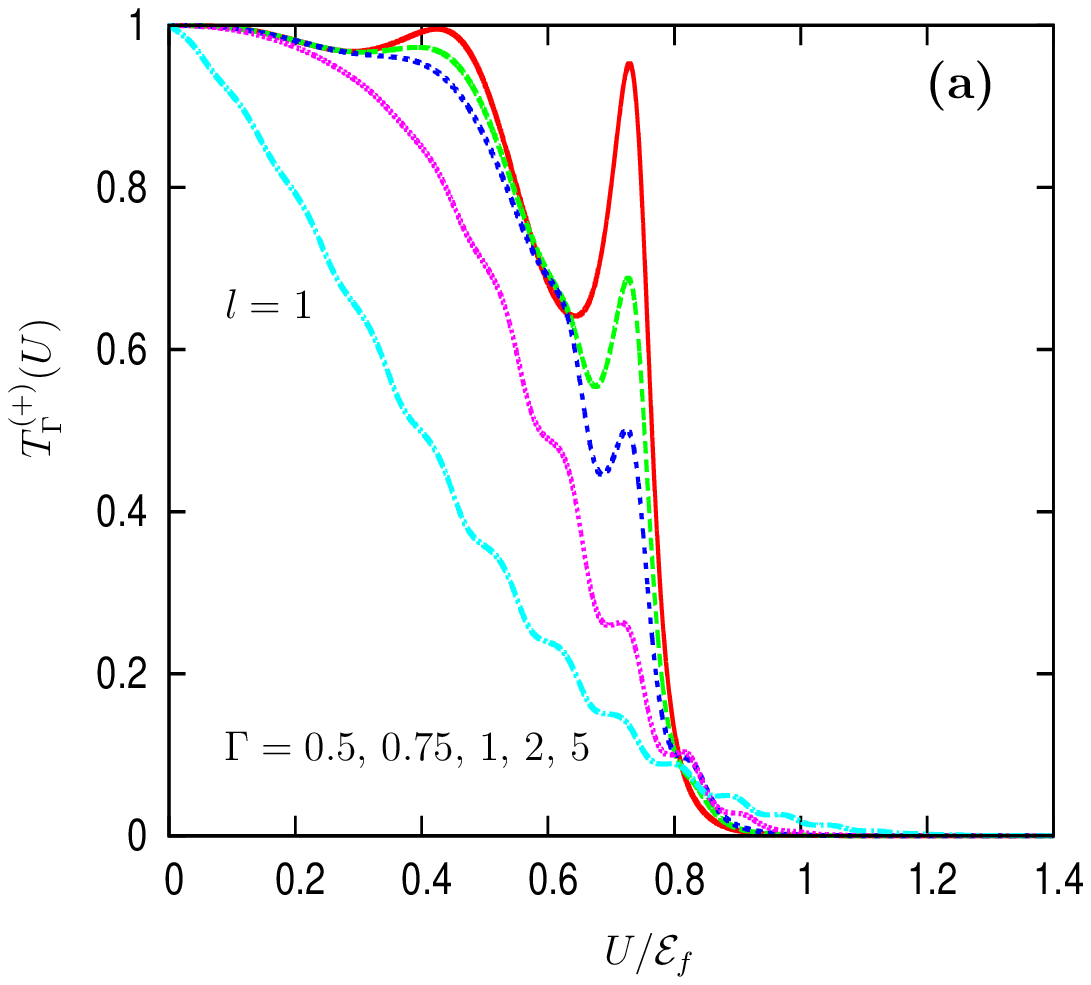}
\includegraphics[width=3.6in]{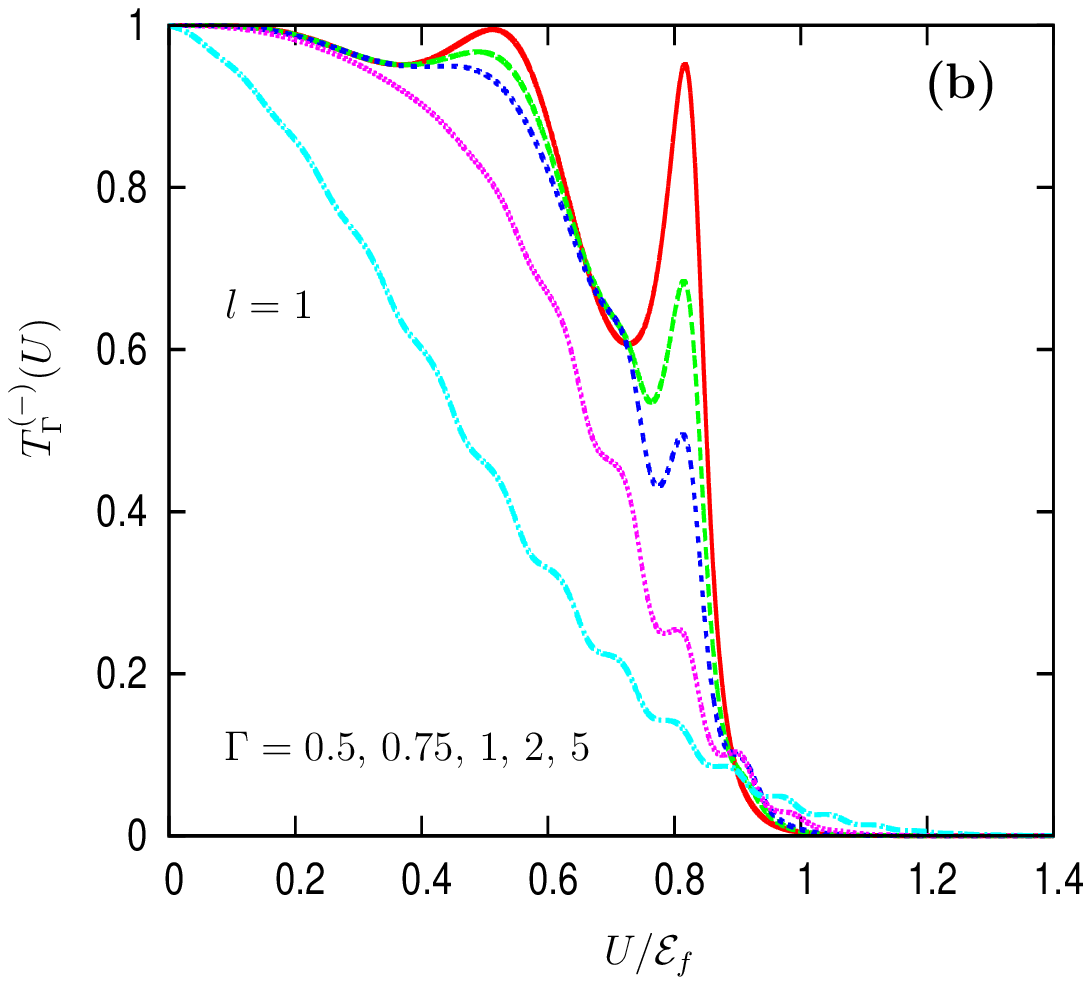}
\caption {(Color online)  Transmission probabilities in the presence of
impurities (defects) included  phenomenologically via Eq. (\ref{broaden}) with the
parameter $\Gamma$ related to temperature for  (a) the $``+$" states
and  (b) the $``-"$ for SOI ${\cal E}_{\alpha}=0.08{\cal E}_f$.}
\label{FIG:6}
\end{center}
\end{figure}
\begin{figure}[p]
\begin{center}
\includegraphics[width=3.6in]{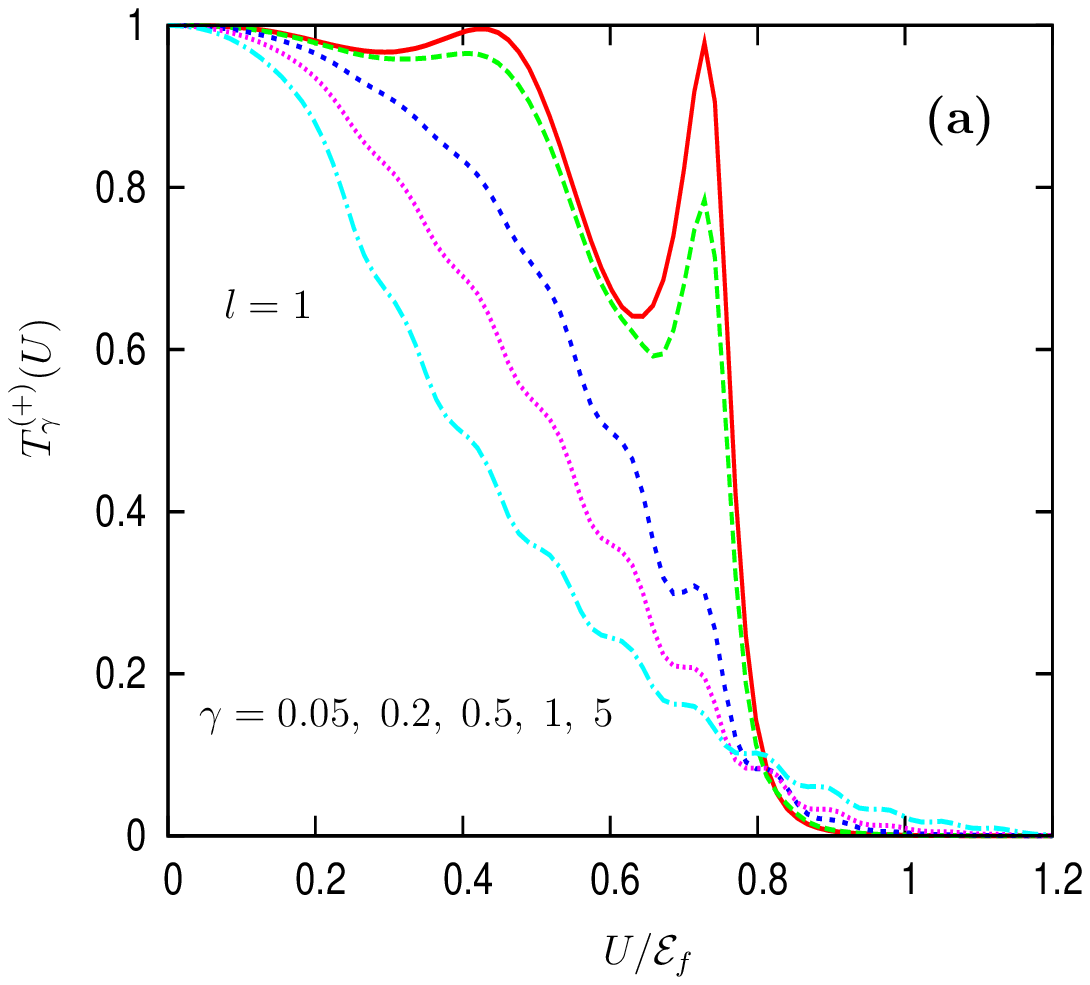}
\includegraphics[width=3.6in]{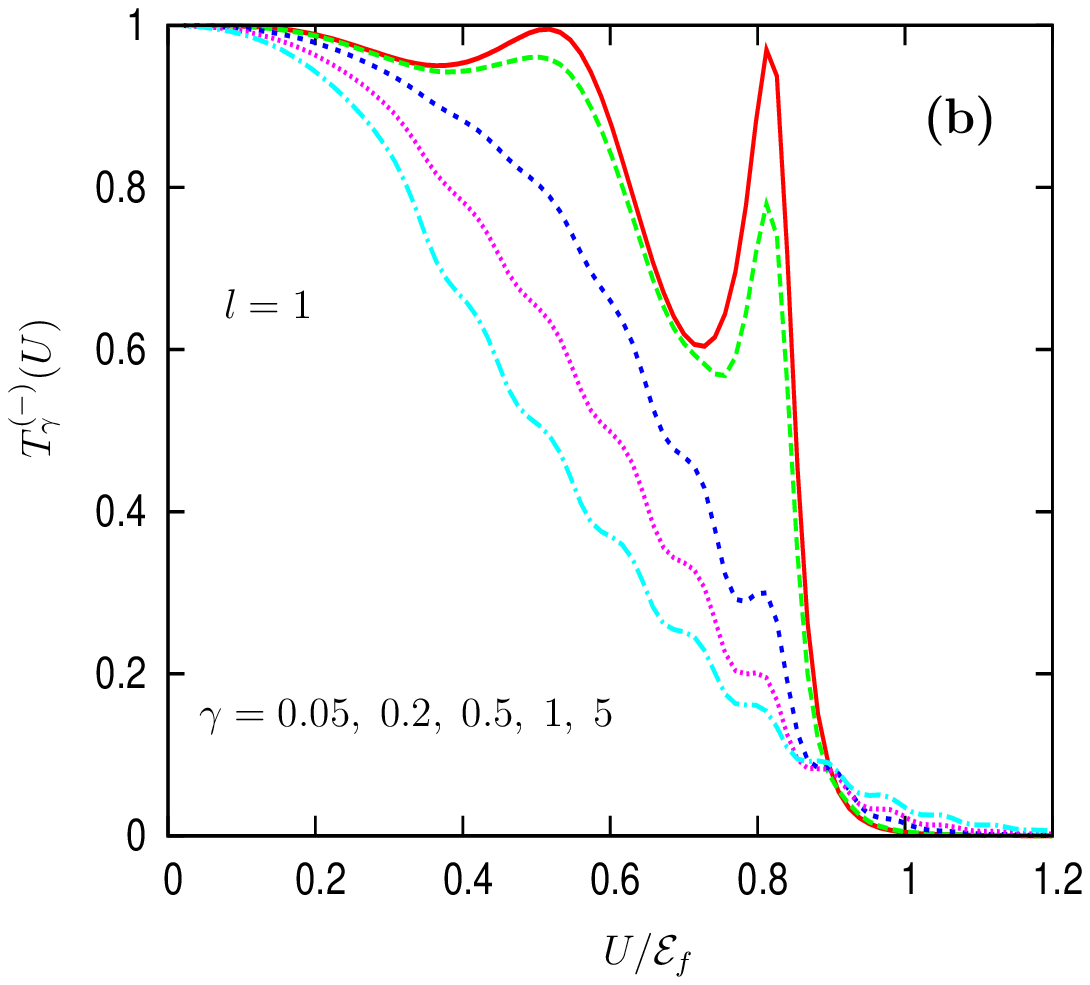}
\caption {(Color online)  Transmission Probabilities in the presence of
impurities (defects) included  phenomenologically via Eq. (\ref{broaden2}) with the
parameter $\gamma$ related to temperature for (a) the $``+$" states
and (b) the $``-"$ for SOI ${\cal E}_{\alpha}=0.08{\cal E}_f$.}
\label{FIG:7}
\end{center}
\end{figure}

In Fig. \ref{FIG:6}, we have plotted  the transmission
probability $T_\Gamma=|t_{s,s^\prime}^{ \rm disorder}(l=1)|^2$
given by Eq.\ (\ref{broaden}) phenomenologically expressing  the ``temperature"
effect for different $\Gamma$s in the absence and presence
of SO coupling.  In both models, the transmission probability is substantially
reduced and there is no perfect transmission at finite barrier height.
Our results show that in the presence of disorder, the SOI has negligible effect on
the transmission probability for low barrier heights. However, the effect of SO
coupling is increased as the barrier height is increased. The effect of disorder
on the transmission probability  $T_\gamma=|t_{s,s^\prime}^{ \rm disorder}(l=1)|^2$
according to Eq. (\ref{broaden2}) for various $\gamma$s is shown in \ref{FIG:7}.
The effect is qualitatively the same as given by Eq. (\ref{broaden}).

\section{Concluding Remarks}
\label{sec5}

In this paper, we assumed that the electrons are confined to move on a single
nanotube. It would be of interest to consider how our results would be affected
when there are two-dimensional electron gases confined to two coaxial tubes in
 the presence of tunneling between the two tubes. We know   that tunneling
 leads to a collective oscillation.\cite{new1}     Electron-electron Coulomb
 interaction gives rise to a shift of the resonance frequency from the particle-hole
 excitation frequencies and to a finite  lifetime of the collective  excitations.
We had shown \cite{newg} that when an ``external"   charged particle travels in the vicinity of
an electron gas on the surface of a nanotube, it gives rise to collective plasmon
excitations of the nanotube due to the frictional force between the electron
gas and the charged particle. The effect of SOI on the energy loss will be
investigated making use of our derived single-particle states.

\acknowledgments

This research was supported by  contract \# FA 9453-07-C-0207 of AFRL.

\newpage

\newpage

\newpage

\newpage

\newpage

\newpage

\end{document}